# THE TWITTER EXPLORER: A FRAMEWORK FOR OBSERVING TWITTER THROUGH INTERACTIVE NETWORKS

Armin Pournaki, Felix Gaisbauer, Sven Banisch and Eckehard Olbrich*

## ABSTRACT

We present an open-source interface for scientists to explore Twitter data through interactive network visualizations. Combining data collection, transformation and visualization in one easily accessible framework, the *twitter explorer* connects distant and close reading of Twitter data through the interactive exploration of interaction networks and semantic networks. By lowering the technological barriers of data-driven research, it aims to attract researchers from various disciplinary backgrounds and facilitates new perspectives in the thriving field of computational social science.

* Max Planck Institute for Mathematics in the Sciences, Leipzig, Germany.

## 1 INTRODUCTION

Due to its public-by-default nature and the possibility of calling data sets conveniently via an API, Twitter has become a widely used source for the observation and analysis of political debates (Conover, Gonçalves, et al. 2011; Gaumont, Panahi, and Chavalarias 2018), sentiments (Paltoglou and Thelwall 2017), brand communication (Nitins and Burgess 2014), or natural disasters (Bruns and Burgess 2014), to name a few. Different kinds of interactions on Twitter (Rainie 2014) are often represented in the form of networks, such as retweet networks (Conover, Gonçalves, et al. 2011; Conover, Ratkiewicz, et al. 2011), reply networks (Gaisbauer et al. 2020), mention networks (Conover, Ratkiewicz, et al. 2011), follower networks (Myers et al. 2014) or co-hashtag networks (Burgess and Matamoros-Fernández 2016). While many of the employed methods, building on concepts from graph theory and network science, can be regarded as distant reading approaches, it is undoubtedly crucial for social science researchers to perform a close reading[1] of digital traces to gain a more focused and specific understanding of their objects of research. As an interface that bridges the two approaches, the *twitter explorer* gives an "overview of the data that highlights potentially interesting patterns", while allowing a "drill down on. these patterns for further exploration" (Jänicke et al. 2015). This means that the structural overview given by the network allows the user to find the relevant content through a framework we present as "guided close reading". In this context, we conceive the *twitter explorer* as a social media observatory, enabling users to "capture the complexities of social behaviour [...] through computational analyses of digital media data" (Willaert et al. 2020).

## 2 PREVIOUS WORK

There exists a wide range of tools for collecting, analyzing and visualizing Twitter data, some of which are referenced on Twitter's own website (Twitter 2020e). Among the most popular tools are DMI tcat (Borra and Rieder 2014) for data collection and analysis in combination with the powerful network visualization suite Gephi (Bastian, Heymann, and Jacomy 2009). While many existing solutions are suited for one specific task and rely on the interplay and compatibility of several applications, the *twitter explorer* provides an open framework that combines data collection, transformation and visualization and allows users to explore the collected Twitter corpus interactively, while being open to external data sources and analysis suites through data import and export. To better situate the *twitter explorer* in its context, a comparison of existing tools is presented in Table 1 below.

[1] These terms were originally coined by Franco Moretti in the context of literary studies (Moretti 2000). Close reading refers to "the thorough interpretation of a text passage" (Jänicke et al. 2015), while distant reading "aims to generate an abstract view by shifting from observing textual content to visualizing global features of a single or of multiple text(s)" (Jänicke et al. 2015).

*Table 1. A comparison of tools for access, analysis and visualization of Twitter data. Due to the steady pace of tool development in this field of research, this list cannot be exhaustive. However, we aim to give an overview of some popular methods and their features. A checkmark in parenthesis denotes basic or experimental functionality. Note that we included almost only open-source software in the table. Furthermore, we chose to omit tools that were not maintained anymore.*

| | **data access** | | **data analysis** | | **data visualization** | | **data flow** | | |
|---|---|---|---|---|---|---|---|---|---|
| | search | stream | statistics | networks | static | interactive | input | output | **last commit** |
| **twitter explorer** | ✓ | – | ✓ | ✓ | ✓ | ✓ | ✓ | ✓ | 1/29/21 |
| **twarc**[2] | ✓ | ✓ | ✓ | ✓ | – | – | – | ✓ | 1/24/21 |
| **DMI tcat**[3] | ✓ | ✓ | ✓ | ✓ | ✓ | (✓) | – | ✓ | 7/20/20 |
| **NodeXL Pro**[4] | ✓ | ✓ | ✓ | ✓ | ✓ | ✓ | ✓ | ✓ | – |
| **Gephi**[5] | – | – | – | – | ✓ | (✓) | ✓ | ✓ | 9/28/20 |
| **Facepager**[6] | ✓ | – | ✓ | ✓ | ✓ | – | – | ✓ | 1/28/21 |
| **Twint**[7] | – | – | ✓ | ✓ | ✓ | – | – | ✓ | 12/17/20 |
| **vosonSML**[8] | ✓ | ✓ | ✓ | ✓ | ✓ | – | ✓ | ✓ | 12/26/20 |
| **SMO-TMAS**[9] | ✓ | ✓ | ✓ | ✓ | ✓ | – | – | – | 11/13/19 |
| **OSoMe**[10] | | | | | | | | | |
| **botslayer/hoaxy** | – | ✓ | ✓ | (✓) | ✓ | (✓) | – | ✓ | 1/12/21 |
| **OSoMe Networks** | – | (✓) | – | – | ✓ | ✓ | – | – | – |

## 3 ARCHITECTURE

The *twitter explorer* consists of three components:

- The collector, a Streamlit-powered[11] (Treuille, Teixeira, and Kelly 2020) application provides a graphical user interface for the Twitter Search API and saves the collected data for further processing.
- The visualizer, a Streamlit-powered application provides a graphical user interface for the generation of interaction networks and semantic networks based on the collected data and saves the interactive networks.
- The explorer interface allows users to interact with the networks and explore the underlying metadata of nodes and links.

Each of these components is conceived in a modular way which facilitates adding new features to the *twitter explorer* (see Figure 1).

[2] DocNow (2020)
[3] Borra and Rieder (2014)
[4] Smith (2013)
[5] Bastian, Heymann and Jacomy (2009)
[6] Jünger and Keyling (2019)
[7] TWINT-Project (2018)
[8] VOSON-Lab (2018)
[9] Young (2020)
[10] Davis et al. (2016)
[11] Streamlit is a Python library for the creation and deployment of data-analytic tools

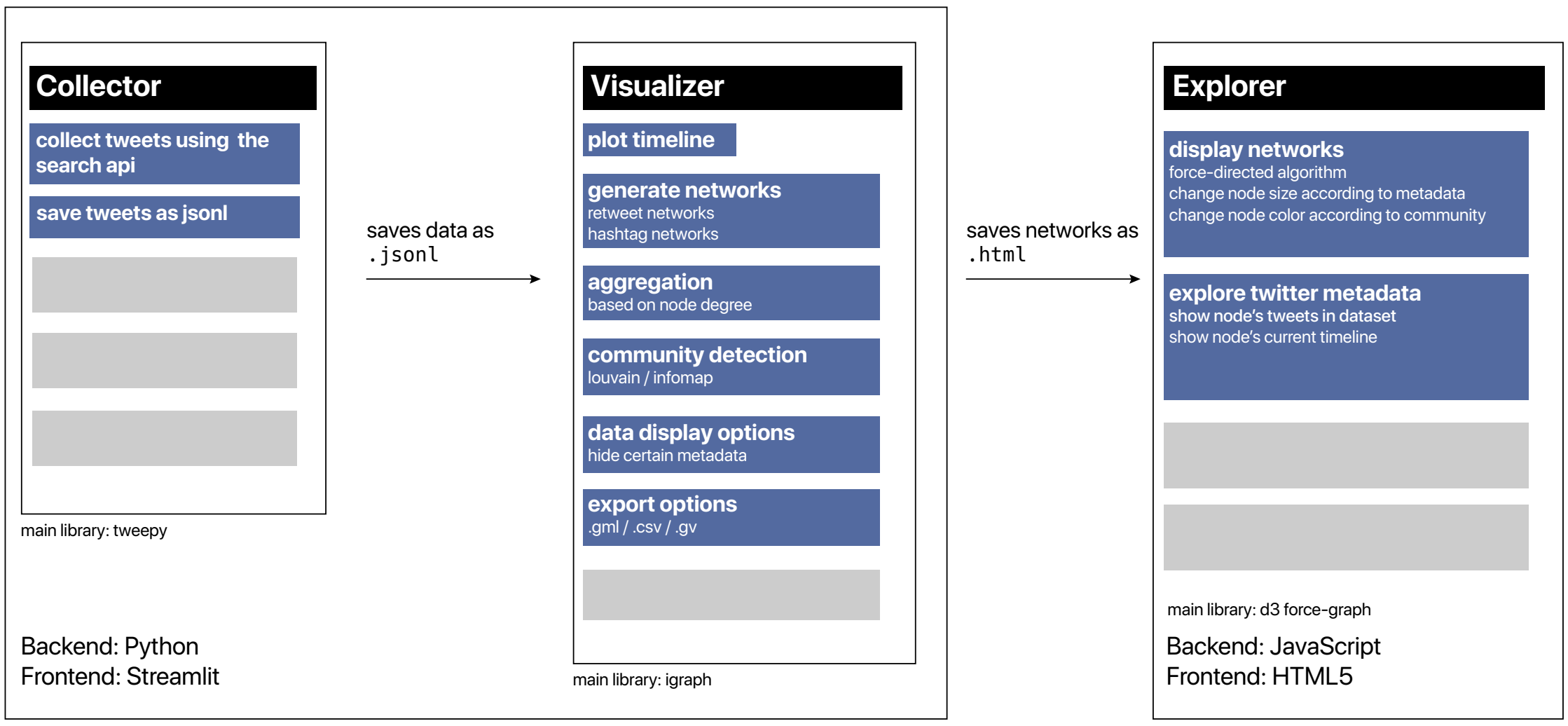

*Figure 1. The twitter explorer framework. The collector (left), after having set up the credentials, allows for connection to the Twitter Search API and saves the collected tweets in jsonl format. They are then passed on to the visualiser (middle), where the user can get an overview of the content and then create the retweet- and hashtag networks. The interactive networks are generated as html files that can be explored in the web browser. The modular structure of the three components facilitates the development of new features, which are suggested by the light grey boxes.*

## 3.1 DATA ACQUISITION: THE COLLECTOR

In the collector, the user interacts with the Twitter Search API (Twitter 2020f), giving access to a limited set of tweets from the last 7 days.

### *3.1.1 Authentication*

Since 2018, users need to apply for a Twitter Developer Account in order to access the API (Roth and Johnson 2018). Since the collector makes direct API calls, this step is necessary for its usage. There are developer accounts specific to academic research (Twitter data for academic research 2020). The user can then create app tokens which will allow the *twitter explorer* to connect to the API via Application-only authentication (OAuth 2.0) (Twitter 2020a).

### *3.1.2 Collection*

There are different APIs for users to collect Twitter data. The Stream API (Twitter 2020g) filters all incoming tweets for a given search string. It can be used to collect tweets containing a certain keyword, or to collect all tweets by a certain (group of) user(s). This API allows the retrieval of all published tweets and is only capped by the upper bound of 1% of the total Twitter traffic. The *twitter explorer* has no built-in feature for the Stream API because we believe that such collections are best done on a headless server which stores the large amounts of incoming data in a database. To collect tweets from the past, we recur to the Search API (Twitter 2020f). The collection of tweets is again initiated by a keyword string, following the rules of a Twitter Advanced Search (Twitter 2020c). This free API comes with limitations: users can only make a limited number of requests per 15 minutes (Twitter 2020d). In the *twitter explorer*, tweets are continuously stored until all possible tweets that the Search API provides are collected.

Note that the Search API gives access only to indexed tweets from the last 7 days. Therefore, a collection created by the Search API cannot be considered extensive, and it is subject to Twitter's nontransparent filtering algorithm. Previous research on the comparison between Stream and Search API however concludes that Twitter filters mostly duplicates and strong language (Thelwall 2015; Black et al. 2012). Measuring the volume of a 48-hour collection of tweets based on the keyword "clubhouse", we find that 80% of tweets from the Stream API collection are contained in the Stream API (see Figure 5 in the Appendix).

## 3.2 DATA TRANSFORMATION: THE VISUALIZER

The visualizer creates interactive network visualizations from the collected corpus. One can distinguish between interaction networks (with users as nodes) and semantic networks (with words or concepts as nodes). The *twitter explorer* currently supports the creation of retweet networks as interaction networks and hashtag co-occurrence networks as semantic networks. Several data aggregation methods allow for exploration of the network at different scales.

### *3.2.1 Twitter timeline*

The data is presented as a timeline, where tweet counts are plotted over time. The user can get a feeling of the overall salience of the chosen keyword and possible peaks can hint towards special events.

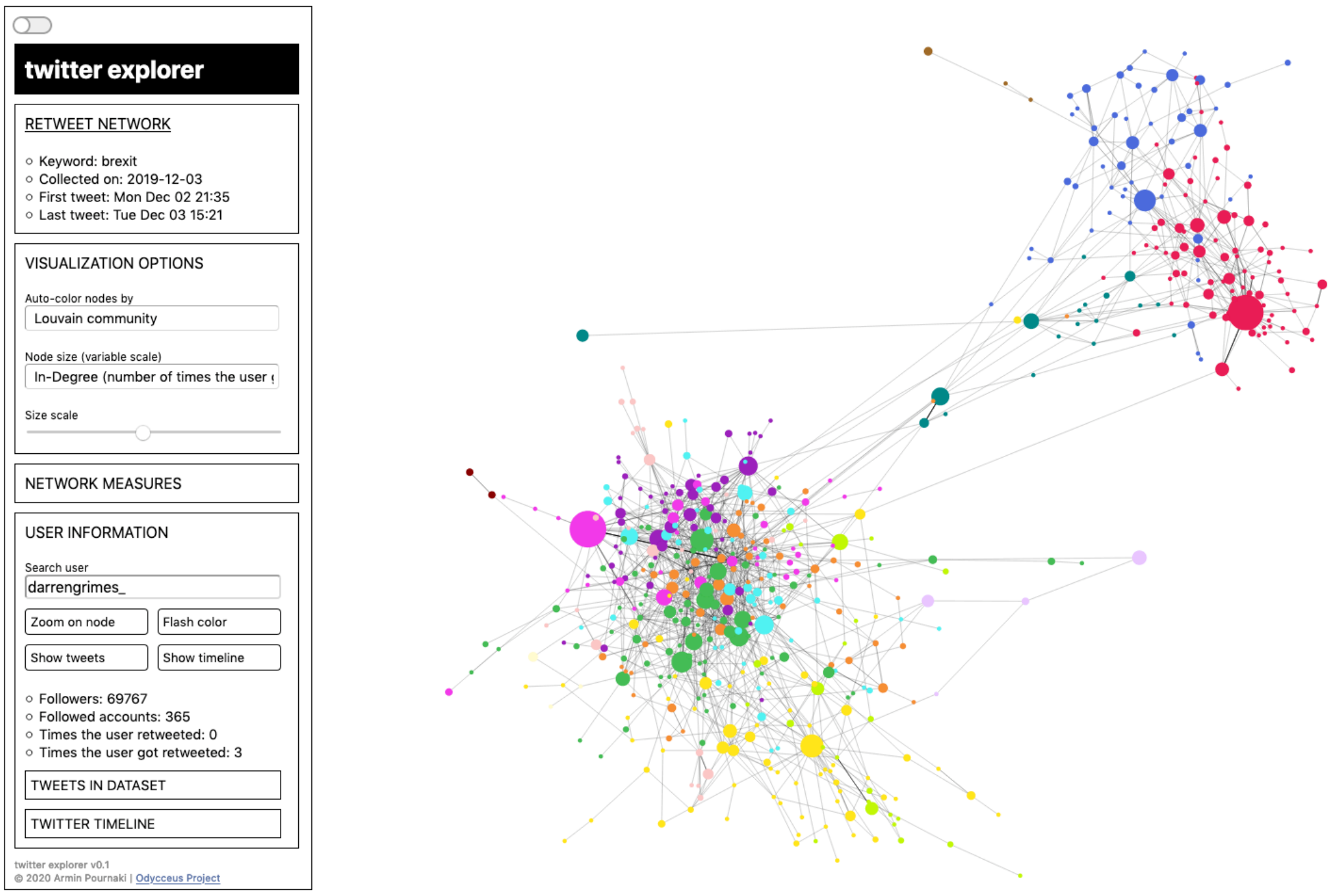

*Figure 2. The retweet network exploration interface. The modular command palette (left) can (1) show information about the underlying data, (2) modify the visualization, (3) display network measures and (4) search for and show information about specific users and the content they generated in the dataset. Nodes are colored according to their community. They can be interacted with by clicking or hovering to display the username and relevant metadata in the palette. We invite the reader to test the interactive visualization here: https://twitterexplorer.org/try.html*

### *3.2.2 Interaction networks*

There are several ways of interaction on Twitter: retweets, mentions, replies, following, likes, quotes and direct messages. Not all of them are accessible through the API. We focus on retweet interaction which can be represented as a directed network in which nodes are users and a link is drawn from node to if retweets . The *twitter explorer*'s visualizer provides an interface for creating retweet networks which includes the following features:

***Community detection.*** In order to find strongly connected clusters of a network, it has become common practice to employ community detection algorithms. The *twitter explorer* currently supports Louvain (Blondel et al. 2008) and InfoMap (Rosvall and Bergstrom 2007) algorithms.

***Force-directed layout.*** The visualization library (Asturiano 2018) spatializes the network using a force-directed layout in which nodes that retweet each other more often are placed closer to each other (Noack 2009).

***Aggregation methods.*** One challenge for understanding and visualizing complex interaction networks is to find useful aggregation methods necessary to observe the underlying discourse at different levels of granularity. We therefore propose several methods of node aggregation: (1) removing nodes that only retweet one source and don't generate any content, (2) removing nodes that were retweeted less than times and (3) reducing the network to an interaction network of communities (cluster graph).

*Hiding sensitive metadata*. Removes all accessible metadata of users that have less than 5000 followers from the interactive visualization. The nodes are visible, and their links are taken into account, but they cannot be personally identified in the interface.

*Export abilities*. Exports the networks to common formats like edgelist, GML or GraphViz. The framework is therefore compatible with a wide range of existing tools for network analysis (Bastian, Heymann, and Jacomy 2009; Peixoto 2014; Csardi and Nepusz 2006).

An example of a retweet network visualized with the *twitter explorer* can be seen in Figure 2. We collected data using the keyword "Brexit" about 10 days before the General Election in the UK in December 2019. We observe a polarized retweet network, where pro and anti-Brexiteers form two distinct clusters. This hints to the fact that users in the debate tend to mainly share (and endorse) content created by their own opinion group.

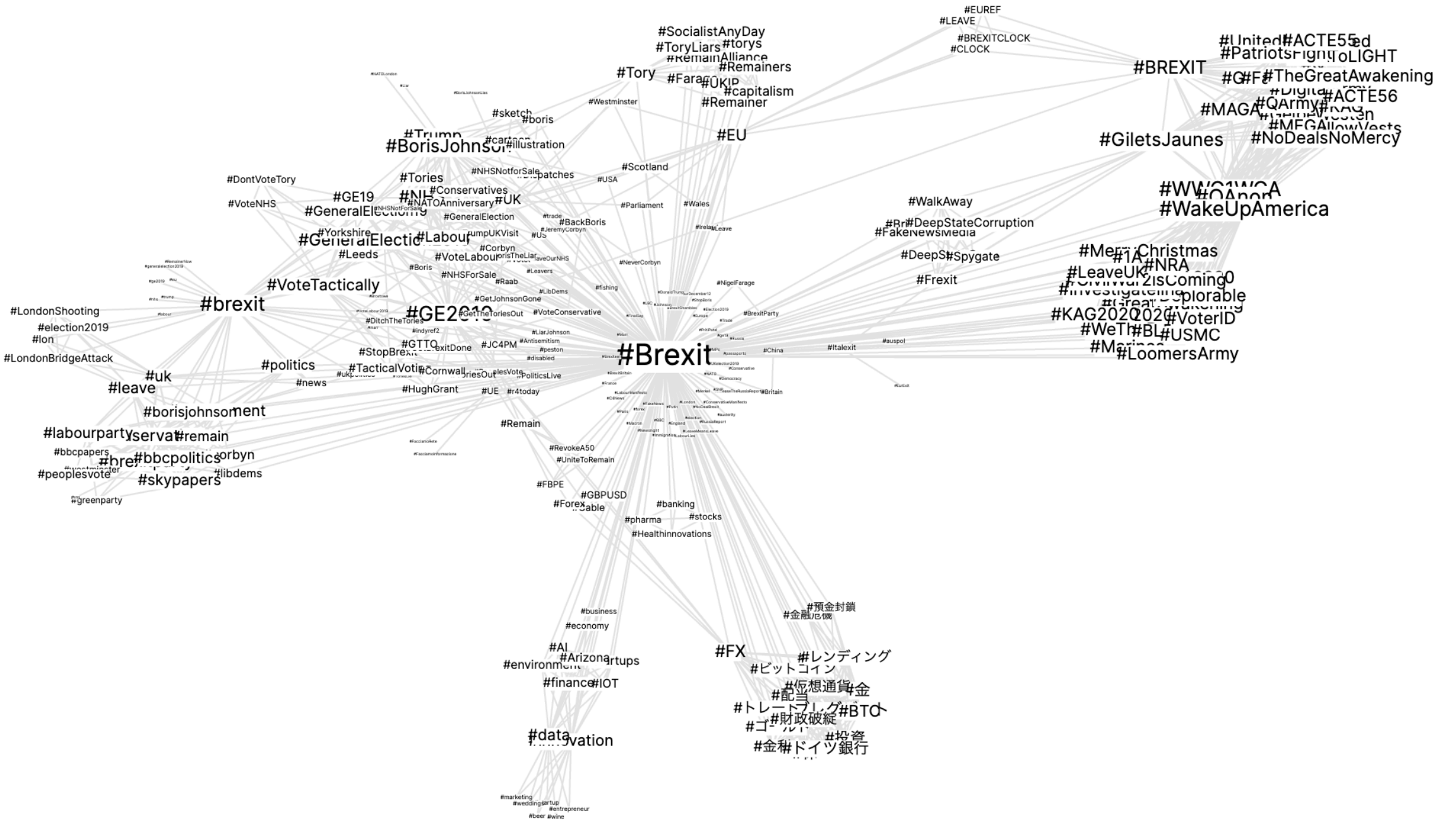

*Figure 3. Hashtag network. Every node is a hashtag, and a link is drawn between hashtags for every tweet they appear in together. The size of the text corresponds is proportional to the node degree. We invite the reader to test the interactive visualization here: https://twitterexplorer.org/try_htn.html*

#### *3.2.3 Semantic networks*

While retweet networks allow to identify the main proponents of a debate and their interaction patterns, looking at the most retweeted tweets might not be sufficient to get an impression of the content structure of the debate. In order to explore the textual content of the data, we propose hashtag co-occurrence networks. Here, every node is a hashtag, and links are drawn between nodes if they appear in the same tweet. By again laying out the network with a force-directed algorithm, the hashtag network gives an overview of the debate's vocabulary and can reveal the different subtopics within a debate.

An example using the previously introduced Brexit data is shown in Figure 3. Hashtags like "#votetactically", "#GetTheToriesOut" or "#VoteConservative" point towards discussions closely related to the General Election, while hashtags like "#DeepStateCorruption", "#TheGreatAwakening" or "#QAnon" shed light on the existence of conspiracy-theory-related sub-discussions in the dataset.

### 3.3 NETWORK EXPLORATION INTERFACE

The *twitter explorer* offers an intuitive exploration interface (see Figure 2). A modular command palette allows for user interaction and provides insight into the underlying meta data of the network:

*Network information*. Accesses generic information about the network (keywords used to collect the data, date of collection, first/last tweet of the dataset).

*Visualization options*. Supports different node colorings according to their community assignment. The node size can be dynamically changed according to their respective metadata values (in/out-degree, number of followers, number of followed accounts). This facilitates for instance the detection of news outlets.

*Network measures*. Shows the number of nodes and links in the network. This set will be extended to include a wider range of network indicators in future releases.

*User information*. Search users in the given network and find them by zooming or flashing their color. Display the user's relevant metadata (number of followers, number of followed accounts, number of retweets, number of times retweeted), their tweets in the dataset as well as their current timeline. Note that the interface will only display tweets that are still online at the time of exploration. By doing so, it complies with the Twitter display requirements (Twitter 2020b).

## 4 INTEGRATION WITH OTHER METHODS

The *twitter explorer* can be regarded an all-in-one-solution for the exploration of Twitter networks, for which it is easy to develop new modules within the existing components (see Figure 1). An example would be to include additional community detection algorithms or new node aggregation methods.

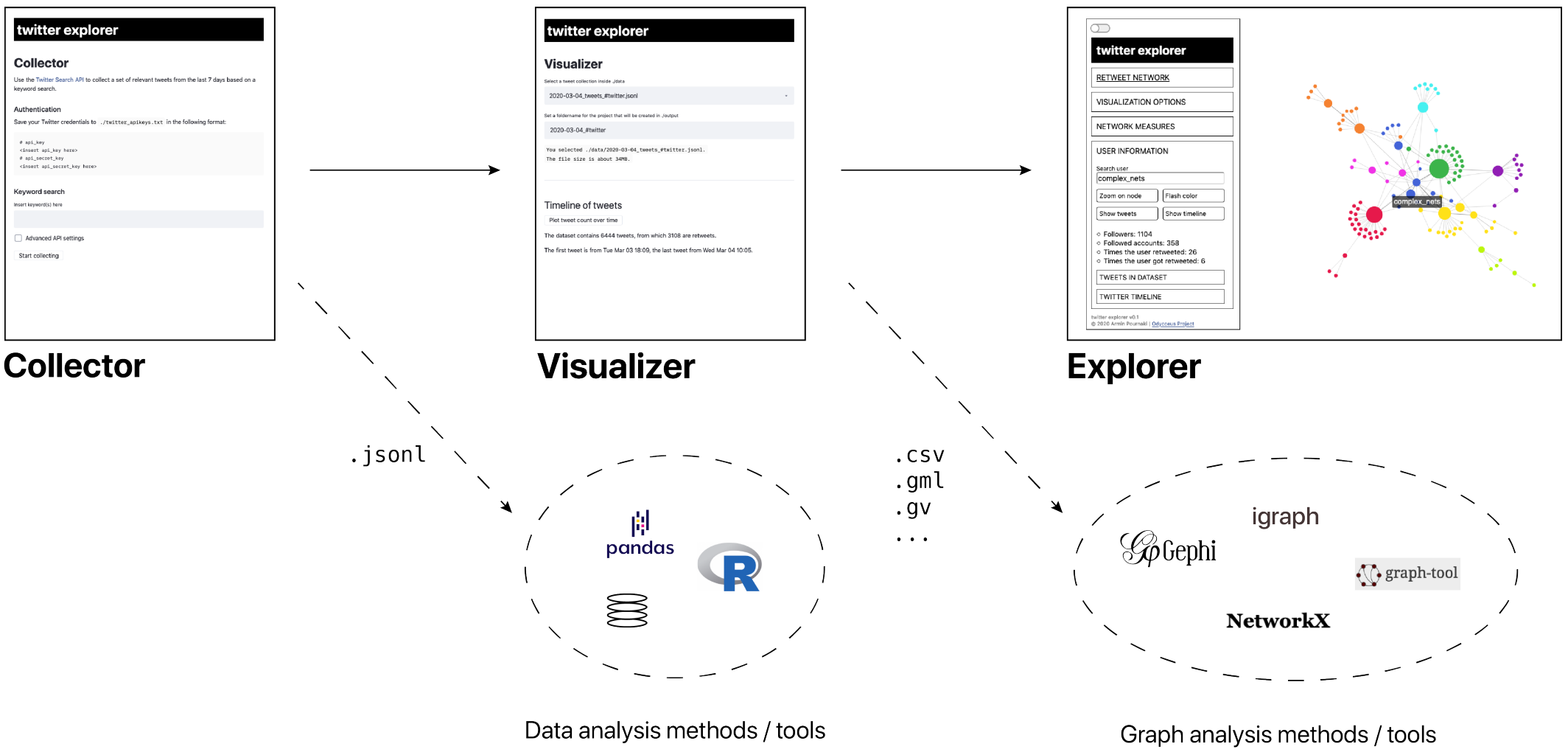

*Figure 4. The twitter explorer in context. Its modular structure makes it easy to develop new features for the twitter explorer, but it also allows it to be used in combination with existing data analysis and network science tools. The dotted arrows depict export paths allowing users to integrate the (transformed) data from the twitter explorer into their desired data analysis environment.*

At the same time, its modular structure (division into collector / visualizer / explorer) and the ability to export the generated data makes the tool compatible with a variety of other data analysis tools (see Figure 4). Therefore, scientists can use the *twitter explorer* in combination with existing tools from data and network science. For instance, after the collector, the data could be passed on to a database, or passed on to a natural language processing pipeline for content analysis. After the visualizer, the exported network can be imported to a visualization suite like Gephi, where various network measures and layout algorithms can be computed.

### 4.1 FUTURE DEVELOPMENT

The *twitter explorer* is currently in an open beta stage on GitHub. Future work will include the dynamical nature of retweet interaction in the visualization paradigms. In order to disseminate the framework and attract new audiences to the field of data-driven research, vignettes (use-cases) will be designed to showcase the *twitter explorer*'s use in social science research. They will be published on our blog which is accessible at https://blog.twitterexplorer.org. Furthermore, it is planned to add the possibility of exploring recently developed measures such as graph curvatures which can provide new insights to the analysis of social networks (Leal et al. 2018). The authors plan to actively maintain the tool and adapt it to Twitter API changes, like the one that was recently announced for Academic Research (Twitter 2021).

## 4.2 AVAILABILITY

The *twitter explorer* interface can be tested at https://twitterexplorer.org. The source code is available on GitHub, where the current release can be downloaded (Pournaki 2020). It is licensed under the GNU GPLv3 license (Free Software Foundation Inc. 2007).

## 4.3 TECHNICAL DETAILS

The *twitter explorer* is written partly in Python (data collection and transformation) and JavaScript (interactive network visualization). The frontend for the data collector and the visualizer is made with Streamlit (Treuille, Teixeira, and Kelly 2020), a Python library for the creation and deployment of data-analytic tools. The Twitter objects are stored in the json lines format (Ward 2020). The network operations and community detection rely on the Python implementation of igraph (Csardi and Nepusz 2006). The interactive networks are drawn using D3.js (Bostock 2011), more specifically the force-graph library (Asturiano 2018).

## AUTHOR CONTRIBUTIONS AND FUNDING STATEMENT

The idea for the *twitter explorer* originated from fruitful discussions in the context of the ODYCCEUS project between Armin Pournaki, Felix Gaisbauer, Sven Banisch and Eckehard Olbrich. The tool is designed and developed by Armin Pournaki. All authors wrote the manuscript. This project has received funding from the European Union's Horizon 2020 research and innovation programme under grant agreement No 732942.

## REFERENCES

Asturiano, Vasco (2018). *force-graph*. https://github.com/vasturiano/force-graph. [Online; accessed 29-January-2021].

Bastian, Mathieu, Sebastien Heymann, and Mathieu Jacomy (2009). "Gephi: An Open Source Software for Exploring and Manipulating Networks". In: url: http://www.aaai.org/ocs/index.php/ICWSM/09/paper/view/154.

Black, Alan et al. (2012). "Twitter zombie: Architecture for capturing, socially transforming and analyzing the Twittersphere". In: *Proceedings of the 17th ACM international conference on Supporting group work*, pp. 229–238.

Blondel, Vincent D et al. (2008). "Fast unfolding of communities in large networks". In: *Journal of statistical mechanics: theory and experiment* 2008.10, P10008.

Borra, Erik and Bernhard Rieder (2014). "Programmed method: Developing a toolset for capturing and analyzing tweets". In: *Aslib Journal of Information Management*.

Bostock, Mike (2011). *D3.js*. https://d3js.org/. [Online; accessed 29-January-2021].

Bruns, Axel and Jean Burgess (2014). "Crisis communication in natural disasters: The Queensland floods and Christchurch earthquakes". In: *Twitter and society [Digital Formations, Volume 89]:* ed. by A Bruns et al. United States of America: Peter Lang Publishing, pp. 373–384.

Burgess, Jean and Ariadna Matamoros-Fernández (2016). "Mapping sociocultural controversies across digital media platforms: One week of# gamergate on Twitter, YouTube, and Tumblr". In: *Communication Research and Practice* 2.1, pp. 79–96.

Conover, Michael D, Bruno Gonçalves, et al. (Oct. 2011). "Predicting the Political Alignment of Twitter Users". In: 2011 IEEE Third International Conference on Privacy, Security, Risk and Trust and 2011 IEEE Third International Conference on Social Computing, pp. 192–199. doi: 10.1109/PASSAT/SocialCom.2011.34.

Conover, Michael D, Jacob Ratkiewicz, et al. (2011). "Political polarization on twitter". In: Fifth international AAAI conference on weblogs and social media.

Csardi, Gabor and Tamas Nepusz (2006). "The igraph software package for complex network research". In: InterJournal Complex Systems, p. 1695. url: http://igraph.org.

Davis, Clayton A et al. (2016). "OSoMe: the IUNI observatory on social media". In: *PeerJ Computer Science* 2, e87.

DocNow (2020). *twarc*. https://github.com/DocNow/twarc. [Online; accessed 29-January-2021].

Free Software Foundation Inc. (2007). *GNU General Public License*. https://www.gnu.org/licenses/gpl-3.0.html. [Online; accessed 29-January-2021].

Gaisbauer, Felix et al. (2020). "How Twitter affects the perception of public opinion: Two case studies". In: *arXiv preprint arXiv:2009.01666*.

Gaumont, Noé, Maziyar Panahi, and David Chavalarias (Sept. 2018). "Reconstruction of the socio-semantic dynamics of political activist Twitter networks—Method and application to the 2017 French presidential election". In: *PLOS ONE* 13.9, pp. 1–38. doi: 10.1371/journal.pone.0201879. url: https://doi.org/10.1371/journal.pone.0201879.

Jänicke, Stefan et al. (2015). “On Close and Distant Reading in Digital Humanities: A Survey and Future Challenges.” In: *EuroVis (STARs)*, pp. 83–103.

Jünger, Jakob and Till Keyling (2019). *Facepager*. https://github.com/strohne/Facepager. [Online; accessed 29-January-2021].

Leal, Wilmer et al. (2018). “Forman-Ricci Curvature for Hypergraphs”. en. In: doi: 10.13140/RG.2.2.27347.84001. url: http://rgdoi.net/10.13140/RG.2.2.27347.84001.

Moretti, Franco (2000). “Conjectures on world literature”. In: *New left review* 1, p. 54.

Myers, Seth A et al. (2014). “Information network or social network? The structure of the Twitter follow graph”. In: *Proceedings of the 23rd International Conference on World Wide Web*, pp. 493–498.

Nitins, Tanya and Jean Burgess (2014). “Twitter, brands, and user engagement”. In: *Twitter and society [Digital Formations, Volume 89]:* ed. by A Bruns et al. United States of America: Peter Lang Publishing, pp. 293–304.

Noack, Andreas (Feb. 2009). “Modularity clustering is force-directed layout”. In: *Physical Review* E 79.2, p. 026102. doi: 10.1103/physreve.79.026102.

Paltoglou, Georgios and Mike Thelwall (2017). “Sensing social media: A range of approaches for sentiment analysis”. In: *Cyberemotions*. Springer, pp. 97–117.

Peixoto, Tiago P. (2014). “The graph-tool python library”. In: figshare. doi: 10.6084/m9.figshare.1164194. url: http://figshare.com/articles/graph_tool/1164194 (visited on 09/10/2014).

Pournaki, Armin (2020). *twitter-explorer*. https://github.com/pournaki/twitter-explorer. [Online; accessed 29-January-2021].

Rainie, Lee (2014). “The six types of Twitter conversations”. In: Pew Research Center 20.

Rosvall, Martin and Carl T Bergstrom (2007). “Maps of information flow reveal community structure in complex networks”. In: arXiv preprint physics.soc-ph/0707.0609.

# APPENDIX

## Stream vs. Search API

We investigate the difference between the Twitter Stream and the Search API. Using the keyword "clubhouse", we first collect tweets using the Stream API from Jan. 25th to Jan. 27th. We then launch the Twitter Search on Jan. 27th to see how many tweets we can collect until Jan. 25th. The tweet count over time is shown in Figure 5. The Search API provides about 80% of the tweets collected by the Stream API. In our example, 13% of the missing tweets in the Search corpus were original tweets and 13% were retweets.

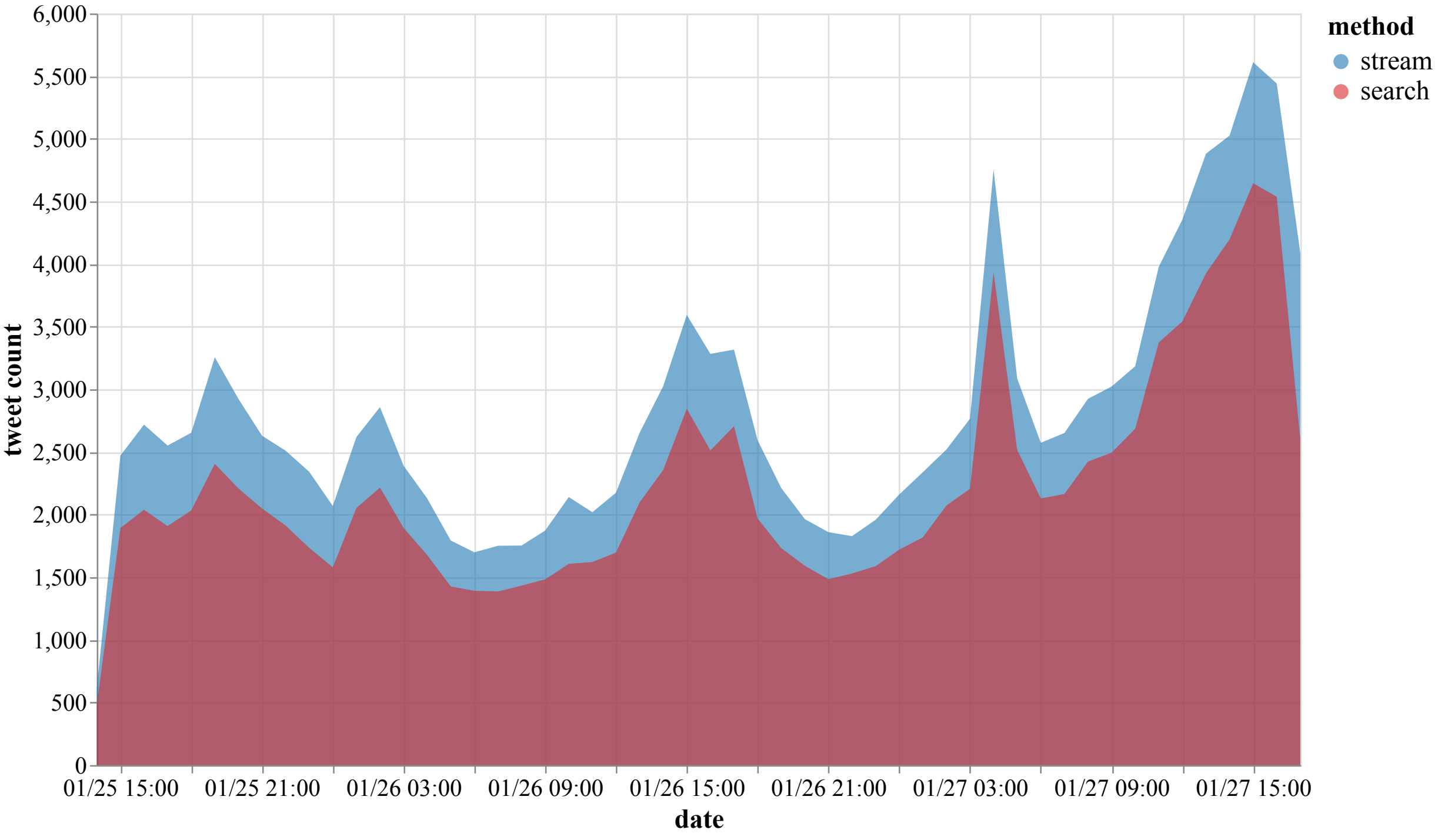

*Figure 5. Streaming API vs Search API. We collected tweets using the keyword "clubhouse" for 48 hours using the search and the streaming API and observe that the Search API constantly returns less tweets than the Search API. Over the whole time range, the searched tweets make out 80% of the streamed tweets.*